\documentclass{emulateapj}

\usepackage{natbib}
\usepackage{epsfig}
\usepackage{appendix}
\usepackage{rotating}
\usepackage{amsmath}
\usepackage{amsfonts} 

\shorttitle{AGN Clustering in the X-ray Band}
\shortauthors{Cappelluti, Allevato \& Finoguenov}

\begin{document}

\title{Clustering of X-ray selected AGN}

\author{
N. Cappelluti\altaffilmark{1,2},
V. Allevato\altaffilmark{3} and
A. Finoguenov \altaffilmark{4,2}}

\altaffiltext{1}{INAF-Osservatorio Astronomico di Bologna, Via Ranzani 1, 40127 Bologna, Italy}
\altaffiltext{2}{University of Maryland, Baltimore County, 1000 Hilltop Circle, Baltimore, MD 21250, USA}
\altaffiltext{3}{Max-Planck-Institut f\"{u}r Plasmaphysik and Excellence Cluster Universe, Boltzmannstrasse 2, D-85748 Garching, Germany}
\altaffiltext{4}{Max-Planck-Institute f\"{u}r Extraterrestrische Physik, Giessenbachstrasse 1, D-85748 Garching, Germany}

\begin{abstract}
The study of the angular and spatial structure of the X-ray sky has been under investigation since the times 
of the {\em Einstein} X-ray Observatory. This  topic has fascinated more than two generations of scientists and slowly 
unveiled an unexpected scenario regarding the consequences of the angular and
spatial distribution of X-ray sources. It was first established from the clustering of sources making the 
CXB  that the source spatial distribution resembles that of  optical QSO. It  then it became evident
that the distribution of X-ray AGN in the Universe was strongly reflecting that of 
Dark Matter. In particular  one of the key result is that X-ray AGN are hosted by Dark Matter Halos
of mass similar to that of galaxy groups.  This result, together with model predictions, has lead to the hypothesis 
that galaxy mergers may constitute the main AGN triggering mechanism. However detailed analysis of observational data, 
acquired with modern telescopes,  and the use of the  new Halo Occupation formalism
has revealed that the triggering  of an AGN
could  also be attributed to   phenomena like tidal disruption or disk instability, and to galaxy evolution. 
This paper reviews results from 1988 to 2011 in the field of X-ray selected AGN clustering. 
\end{abstract} 


\section{Introduction} 

%

After about 50 years from the opening of the X-ray window on the Universe with the discovery 
of Sco-X1 and the Cosmic X-ray background (CXB, Giacconi et al. 1962), our knowledge
of high energy processes in the Universe has dramatically improved.    One of the leading mechanism for 
the production of X-ray in the Universe is accretion onto  compact objects. For this reason the study of 
astrophysical X-ray sources is a powerful tool for studying matter under the effects of extreme gravity. 
As the efficiency of converting matter into energy in accretion processes is proportional to 
the "compactness" of the object, (i.e. $\propto$ M/R), it is clear that the strongest sources
powered by accretion are Super-Massive Black Holes (SMBH).
It also became a cornerstone of astrophysics that every galaxy with a bulge-like component
hosts a SMBH at its centre and that the BH mass and the bulge velocity dispersion  are strictly related   \citep{Mag98}.
  It is also believed that black holes reach those high masses 
via one or more phases of intense accretion activity and therefore shining as  Active Galactic Nuclei (AGN). 
It is believed that an AGN is basically shine mostly from the power emitted 
by a thin, viscous,  accretion disk orbiting the central SMBH \citet{shakira}.   
Such a  disk  produces a high amount X-rays  both from its  hot inner regions (as far as the soft X-ray
emission is concerned)
 and from a non thermal  source which is supposed to be the primary source of 
 X-rays (both soft and hard). 
  
Since its discovery, the nature of the CXB has been strongly debated, but soon 
the community converged into interpreting most of the CXB as the integrated emission of 
AGN across the cosmic time. While the discrete nature of the CXB
has been proposed \citep{b67} and rapidly unveiled by experiments like
 {\em Einstein} \citep{g79} and ROSAT \citep[see e.g.][]{has93},
little cosmological information has been obtained from samples of
AGN because of the scarce number of detected sources 
in the X-ray band.  
Structure formation models and numerical simulations 
have shown that structures in the Universe have undergone a hierarchical
growth starting from the denser peaks   in the primordial gaussian matter distribution.
 The Large Scale Structures (LSS) of the Universe are gravitationally 
dominated by Dark Matter (DM) and we can consider it  as the responsible
and one of the main driver of the Cosmological structures evolution. Dark matter is believed to clump 
in large scale halos \citep[DMH][]{Nav97} which are populated by galaxies. 
Thus galaxies can be considered as tracers of the DM distribution in the Universe
and the study of their spatial clustering led us to a most comprehensive view of the LSS. 
On the other hand AGN/Quasar, as phase of the galactic evolution, is
a quite rare phenomenon in the Universe as their space density of these objects is about 1/100-1/1000 lower than that of galaxies. 
This means that AGN/Quasar survey require large field of view and/or deep exposure to provide statistically 
significant samples.\\
The study of their clustering and its evolution is a powerful tool to understand, from a statistical point 
of view, what kind of environment is more likely to host AGN. This is not just an academic question but,
this is strictly related to the mechanism of AGN activation. We know that one of the  candidate mechanism 
for triggering an AGN is galaxy merger \citep[see e.g.][]{Hop07,Hop08,Hop09,Sil11}. The probability of such an event is definitely
dependent on the environment inhabited by the host galaxy. Even if the mean distance between galaxies is relatively
small, in high density (mass) environments, they have a high velocity dispersion and therefore, 
the likelihood of a major merger is very low.  On the contrary, in the field  the likelihood of galaxy mergers 
is low because of the large average distance between galaxies. 
The most favorable place to 
detect a merger is therefore a moderately low density (mass) environment like a group \citep[see e.g][]{apple}. \\
In fact, merger-driven models \citep[see e.g.][]{Hop07} accurately predict the observed large-scale clustering
of quasars as a function of redshift up to z$\sim$ 4. The clustering is precisely 
that predicted for small group halos in which major mergers of gas-rich galaxies should proceed most efficiently.
Thus it is well established empirically and with theoretical predictions that quasar
clustering traces a characteristic host halo mass $\sim4 \times 10^{12}h^{-1}M_{\odot}$,
supporting the scenario in which major mergers dominate the bright quasar populations.

In addition other phenomena like secular processes may become dominant at lower luminosities as 
suggested by \citet{milo06,Hop06,Hop09}. Low-luminosity AGN could be triggered in more 
common nonmerger events, like stochastic encounters of the 
black holes and molecular clouds, tidal disruption or disk instability.
This leads to the expectation of a characteristic transition to merger-induced
fueling around the traditional quasar-Seyfert luminosity division (growth of BH masses
above/below $\sim 10^{7}M_{\odot}$).
However the triggering mechanism of the SMBH growth must be compliant with M$_{BH}$-$\sigma$ relation,
that links the growth of the SMBH with  growth of the bulge of the host galaxy \citep{Mag98}.\\
As shown in \citet{Hop08}, the predicted large-scale bias of quasars triggered by secular processes
is, at all redshifts, lower than the bias estimated for quasars fueled
by major mergers. This implies that low-luminosity Seyfert galaxies live in DMHs that
never rich the characteristic mass associated with small group scales. \\
On the other hand, the majority of the results on the clustering of X-ray selected AGN, 
suggest a picture where moderate-luminosity AGN live in massive DMHs 
(12.5$<logM_{DMH}$ $[h^{-1}M_{\odot}]<$13.5) up to $z\sim 2$, i.e. 
X-ray selected AGN samples appear to cluster more strongly than bright quasars. 
The reason for this is not completely clear but several studies argued that these large bias
and DMH masses could suggest a different AGN triggering mechanism respect to bright quasars 
characterized by galaxy merger-induced fueling.

This paper reviews results of clustering of X-ray selected AGN from the first {\em Einstein} to the most
recent {\em Chandra} and XMM-{\em Newton} surveys. We 
give a detailed description of the methods used in this kind of analysis from 
simple power-law to halo models. In addition we discuss the results of X-ray AGN 
clustering in the framework of AGN evolution and triggering. 
We adopt a $\Lambda$CDM Cosmology with 
$\Omega_{\Lambda}$=0.7, $\Omega_{m}$=0.3, H$_0$=100 h$^{-1}$ km/s/Mpc 
with h=0.7 and $\sigma_8$=0.8 \citep[][WMAP-7]{lar11}.

\section{Previous measures of X-ray clustering amplitude}
As far as the X-ray source clustering results are concerned, the development of the field
has always be driven by with the performance of the telescopes. In particular while first results 
studied the angular distribution of the unresolved CXB under the assumption that Quasars were its main
contributors, recent  {\em Chandra} and XMM-{\em Newton} surveys sample clustering of AGN with 
a precision comparable to that achievable with redshift galaxy surveys. \\ 
In the following section we will use the following convention for reporting results of 
clustering analysis in the case of power-law representation of the auto(cross)-correlation function: 
if the clustering is measured in the angular space,  we will use: 
\begin{equation} 
w(\theta)={\theta/\theta_0}^{1-\gamma},
\end{equation}
where $\theta_0$ is the angular correlation length.
If the measurements has been performed in the real (redshift) space this becomes:
 \begin{equation} 
\xi(r)={r/r_0}^{-\gamma}~~~~(\xi(s)={s/s_0}^{-\gamma}, in~z-space),
\label{pl}
\end{equation} 
where $\gamma$ is the 3D correlation slope and r$_0$ or s$_0$ are the
correlation lengths.
 \citet{xbf} measured with {\em Einstein} a clustering signal of the CXB on scales $\leq5^{\arcmin}$ corresponding
to an angular correlation length
 $\theta_{0}\sim$ 4$^{\arcmin}$.  
They have shown the  importance of studying the angular structure of the CXB  by  pointing out 
that a large fraction of the CXB could have been attributed  to sources 
with a redshift distribution similar to optical QSOs. In additon,
the first prediction was not consistent with the hypothesis that the 
CXB was also partly produced by a diffuse hot Intergalactic Medium (IGM) component.
It was also proposed that these sources were actually clustered on comoving scales of the order of $\sim$10 $h^{-1}$ Mpc. \\
 \citet{c92}, \citet{g93} and \citet{sh94} observed that the CXB was highly isotropic on scales of the order of 2$^{\circ}$-25$^{\circ}$.
The first attempt of measuring the clustering of X-ray selected AGN was performed by \citet{b93}, that measured a barely significant
signal by using a sample of 183 EMSS sources, mostly local AGN (z$<$0.2).
These evidences have brought the attention to the study of the clustering of the CXB down to the arcminute scale. 
The first significant upward turn for the measurement of AGN clustering in the X-ray band has been brought to light
by ROSAT. By using a set of ROSAT-PSPC pointing on an area of $\sim$40 $deg^{2}$, \citet{v95} measured, for the first time 
an angular correlation signal of faint (ROSAT) X-ray sources on scales $<$10$\arcmin$. By using the Limber equation \citep[see Appendix and][]{pee80}  they have de-projected
their angular correlation function into a real space correlation function and found that, under the assumption that the the redshift 
distribution of the sources was the same as that of optical QSOs, the spatial correlation length 
was in the  range 6-10 h$^{-1}$ Mpc. With such a result, they confirmed the hypothesis that
the CXB was mostly produced by sources with a redshift distribution comparable to that of optically selected QSO, though with almost double source
density.  By using the results of \citet{v95} and \citet[][who  obtained similar results]{a00}, \citet{b01} has shown for the first time that X-ray selected AGN are highly biased tracers of 
the underlying LSS at z$<$1 by showing a redshift evolving bias factor  as large as b$\sim$2. 

However, it is worth to  consider that the deprojection of the angular correlation function into a 3D correlation
 relies on several assumptions, like the model dependent expected redshift distribution, which may  lead 
 to a biased estimate of the real space clustering.  It is however worth noticing
 that angular correlation can be very useful to provide a first
  overview in the early phase of surveys, when optical identifications are not available,
  especially sampling   new part of the parameter space of sources, like i.e.  new unexplored luminosity/flux  limits and 
  therefore source classes. Detailed physical models are however much better investigated by more sophisticated 
  techniques as shown in the following parts. 

The first firm detection of  3-D spatial clustering of X-ray selected AGN has been claimed  by \citet{Mul04} by using data
of the ROSAT-NEP survey. They detected on an area of $\sim$81 deg$^{2}$ a 3$\sigma$ significant signal in the redshift space auto-correlation function of
soft X-ray selected sources at $<z>\sim$ 0.22. They have shown that at that redshift AGN cluster with a typical correlation length r$_{0}$=7.4$\pm{1.9}$ h$^{-1}$ Mpc.
Their results suggest that the population of AGN in such a sample is consistent with an unbiased population with respect to the 
underlying matter . Their result
suggested that at that redshift AGN were hosted in DMHs of mass of the order of 10$^{13}~h^{-1}$ M$_{\odot}$.

With the development of {\em Chandra} and XMM-{\em Newton} surveys and thanks to the high source surface densities
(i.e. $>$400-1000 deg$^{-2}$) our capabilities in tracing the LSS has dramatically 
increased. One of the first evidences that AGN are highly correlated with the underlying LSS has been pointed out 
by \citet{Cap01} and \citet{Cap05} and references therein, who showed
that around massive high-z galaxy clusters the source surface density of {\em Chandra} point sources is significantly, up to two times,
higher than that of  the background.  More recently, \citet{kou10} showed that although the X-ray source surface density
of AGN around galaxy clusters is larger than in the background, the amplitude of their overdensities is about 4 times  
lower than that of galaxies in the same fields. This has been interpreted as a clear indication of an environmental 
influence on the AGN activity. Silverman et al. (2011) in the COSMOS field and \citet{Kos10} in the {\em Swift}-BAT all-sky survey 
have shown  that the AGN fraction in galaxy pairs is higher  relative to isolated 
galaxies of similar stellar mass providing an additional evidence of the influence of the environment on AGN activity.

 {\em Chandra} and XMM-{\em Newton} performed several blanck sky extragalactic surveys,  and most of them dedicated part of their 
 efforts in the study of the LSS traced by AGN to unveil their co-evolution. 
 \citet{Bas04,Bas05} by using data of the XMM-{\em Newton} 2dF-survey have measured an unexpected
 high correlation length both in the angular ($\theta_{0}\sim$10$^{\arcmin\arcmin}$)  and, by projection, in 
 the real space (r$_{0}\sim$16 h$^{-1}$ Mpc). Such an high correlation length has been detected in this field only, 
 thus one can explain such a measurement as a statistical fluctuation.  With the same technique, \citet{Gan06}
 obtained a marginal 2-3$\sigma$ detection of angular clustering in the XMM-LSS survey and obtained $\theta_{0}$=6.3(42)$\pm{3}(^{+7}_{-13}$)
 in the 0.5-2 (2-10) keV bands and a slope $\gamma\sim$2.2. 
 \citet{Puc06} measured the clustering of X-ray sources in the XMM-{\em Newton} ELAIS-S1  
 survey in the soft and hard energy bands with a sample of 448 sources. 
 They obtained $\theta_{0}$= 5.2$\pm{3.8}$ 4$^{\arcmin\arcmin}$ and $\theta_{0}$=12.8$\pm{7.8}$ 4$^{\arcmin\arcmin}$ in the two bands respectively.
 These measurements have been deprojected with the Limber's inversion in the real space
 and obtained r$_{0}$=9.8-12.8 h$^{-1}$ Mpc and r$_{0}$=13.4-17.9 h$^{-1}$ Mpc in the two bands, respectively. 
  
 In the {\em Chandra} era,  \citet{Gil05} measured the real space auto-correlation 
function of point sources in the CDFS-CDFN. They have measured in the CDFS  r$_{0}$=8.6$\pm{1.2}$
 h$^{-1}$ Mpc at z=0.73, while in the CDFN they obtained r$_{0}$=4.2$\pm{0.4}$ h$^{-1}$ Mpc. 
  The discrepancy of these measurements
 has been explained with variance introduced by the relatively
 small field of view and the consequent random sampling of 
 LSSs in the field. 
 In the CLASXS survey \citet{Yan06} obtained a measurement of the clustering 
 at z=0.94 with r$_{0}$=8.1$^{+1.2}_{-2.2}$ h$^{-1}$ Mpc which proposes that AGN are hosted by 
 DMH of mass of 10$^{12.1}$ h$^{-1}$ M$_{\odot}$ (see next Section).  In addition they proposed 
 that AGN clustering evolves with  luminosity and  they 
 found that the bias factor evolves with the redshift. Such a behavior is  similar to that found in optically selected quasars.
 The XMM-{\it Newton} \citep{Has07,Cap07,Cap09} and {\em Chandra} \citep{Elv07,Puc09} survey of the COSMOS field have provided a leap forward to the 
 field of X-ray AGN clustering by surveying a 2 deg$^{2}$ field of view. The key of the success of this project is a redshift survey {\it zCOSMOS} \citep{Lil07}  
 performed simultaneously with the  X-ray survey, together with  observations in more than 30 energy bands from radio to X-ray,
 that allowed to measure either the spectroscopic or the photometric redshift of every source. 
 In the X-ray band, the survey covers 2 deg$^{2}$ with XMM-{\em Newton} with a depth of $\sim$60 ks with the addition of  a central 0.9 deg$^{2}$ 
 observed by {\it Chandra} with $\sim$150 ks exposure. The first sample of $\sim$1500 X-ray sources \citep{Cap07} 
 has been used by \citet{Miy07} to determine their angular correlation function, 
 without knowing their distance, and just assuming a theoretical redshift distribution for the purpose 
 of Limber's deprojection.  
 Significant positive signals have been detected in the 0.5-2 band, in the angular range of 0.5$^{\arcmin}$-24$^{\arcmin}$, 
 while the positive signals were at the $\sim$ 2$\sigma$ and 3$\sigma$ levels in the 2-4.5 and 4.5-10 keV bands, respectively. 
 With power-law fits to the ACFs without the integral constraint term, they have found   correlation lengths of 
$\theta_0$=1.9 $\pm{0.3} ^{\arcmin\arcmin}$, 0.8 $^{+0.5}_{-0.4}$ $^{\arcmin\arcmin}$, 
and 6 $\pm{2}$ $^{\arcmin\arcmin}$ for the three
 bands, respectively, for a fixed slope $\gamma$=1.8. The inferred comoving
  correlation lengths were r$_{0}$=9.8$\pm{0.7}$, 5.8$^{+1.4}_{-1.7}$, and 12$\pm{2}$ h$^{-1}$ Mpc at the effective redshifts of z=1.1, 0.9, and 0.6, 
 respectively.  Comparing the inferred rms fluctuations of the spatial distribution of AGNs $\sigma_{8,AGN}$ (see Appendix)  with those of the underlying dark matter, 
 the bias parameters of the X-ray source clustering at these effective redshifts were found in the range b = 1.5-4. Such a result lead to the 
  conclusion that the typical mass of the DMH hosting an AGN is of the order  M$_{DMH}\sim$10$^{13}$ M$_{\odot}$ h$^{-1}$. 
Similar results have been found by \citet{Ebr09} using the angular correlation function of 30000 X-ray sources
in the AXIS survey.
In the XMM-LSS survey \citet{eli}  measured the clustering of $\sim$5000 AGN and computed via Limber's deprojection
the obtained r$_0$= 7.2$\pm{0.8}$ Mpc/h and  r$_0$= 10.1$\pm{0.8}$ Mpc/h and $\gamma\sim$2 in the 0.5-2 keV and 2-10 keV
energy bands, respectively.
  In the XMM-COSMOS field \citet{Gil09} measured the clustering of 562 X-ray selected and spectroscopically confirmed
  AGN. They have obtained that the correlation length of these source,  r$_0$=8.6$\pm{0.5}$ h$^{-1}$ Mpc and slope of $\gamma$=1.88$\pm{0.07}$.
  They also found  that if source in redshift spikes are removed the correlation length decreases to about 5-6 h$^{-1}$ Mpc . 
  Even if not conclusively,  they also showed that narrow line AGN and broad line AGN cluster in the same way, indicating that both class
  of sources share the same environment, an argument in favor of the unified AGN model which predicts that obscuration,
 and therefore the Type-I/Type II dichotomy is simply a geometrical problem. 
 However it is worth noticing that such a procedure may artificially 
  reduce the clustering signal and the effects of such a cut in the sample, may lead to an unreliable 
  estimate of the clustering signal.
  
  Even if the results of \citet{Gil09} provide a quite complete overview of the environments of the AGN 
  in the COSMOS field, \citet{Alle11} analyzed the same field by using the halo model formalism (see Section \ref{sec1}).
Their results show that  AGN selected in the 
X-ray band are more biased than the more luminous optically selected QSO. 
This observation  significantly deviates
from the prediction of models of merger driven AGN activity \citep{Hop06,Bon09}, indicating that other mechanisms like
disk/bar instability of tidal disruptions may trigger an AGN. 
They also found that Type 1 AGN are more biased than Type 2 AGN up to redshift of $\sim$ 1.5.\\
In the B\"ootes field \citet{Hic09} explored the connection between different classes of AGN
and the evolution of their host galaxies, by deriving
host galaxy properties, clustering, and Eddington ratios of AGN selected in the
radio, X-ray, and infrared (IR) wavebands from the wide-field (9 deg$^2$) B\"ootes
survey. They noticed that radio and X-ray
AGNs reside in relatively large DMHs (M$_{DMH}\sim$3$\times$10$^{13}$ and 10$^{13}$
M$_{\odot}$ h$^{-1}$, respectively) and are found in galaxies with red and green colors. In
contrast, IR AGN are in less luminous galaxies, have higher Eddington ratios,
and reside in halos with M$_{DMH}<$10$^{12}$ M$_{\odot}$ h$^{-1}$. 

On the same line, \citet{Coi09} measured the clustering of 
non-quasar X-ray active galactic nuclei at z = 0.7-1.4 in the AEGIS field. 
Using the cross-correlation of Chandra-selected AGN with ~5000 DEEP2 galaxies they measured a 
correlation  length of r$_{0}$ = 5.95$\pm{0.90}$ h$^{ -1}$ Mpc and slope $\gamma$= 1.66$\pm{0.22}$.
They also concluded that X-ray AGN have a similar clustering amplitude as red, quiescent and "green" 
transition galaxies at z $\sim$1 and are significantly more clustered than blue, star-forming galaxies.
In addition they proposed  a "sequence" of  X-ray AGN clustering, where its strength is primarily 
determined by the host galaxy color; AGN in red host galaxies are significantly 
more clustered than AGN in blue host galaxies, with a relative bias that is similar to that of red to blue 
DEEP2 galaxies. They did not observe any dependence of clustering on optical brightness, X-ray luminosity, 
or hardness ratio. In addition they obtained evidence that galaxies hosting X-ray AGN are more likely to reside 
in groups and more massive DMHs than galaxies of the same color and luminosity without an X-ray AGN. 
 \citet{Alle11}, \citet{Coi09, Mou11} concluded that DEEP2 X-ray AGN at z $\sim$ 1 are more 
clustered than optically selected quasars (with a 2.6$\sigma$ significance) and therefore may reside in more massive DMHs.
In an evolutionary picture their results are consistent with galaxies undergoing a quasar phase while in the blue cloud before settling 
on the red sequence with a lower-luminosity X-ray AGN, if they are similar objects at different evolutionary stages \citep{Hic09}. 
At lower redshift, \citet{Kru10} confirmed the results of \citet{Coi09}.
 Various recent works have been presented indications and/or evidences, of varying significance, regarding a 
correlation between the X-ray Luminosity and the AGN clustering amplitude, based either on the spatial \citep{Yan06,Gil09, Coi09, Cap10, Kru10, Kru11}, 
or the angular  \citep{Pli08} correlation function. \\
Note that luminosity dependent clustering is one of the key features of merger triggered AGN activity and is one of the
prime motivations for AGN clustering analyses. Low L$_X$ AGN have been found to cluster in a similar way 
as blue star forming galaxies while high L$_X$ AGN cluster like red passive galaxies. 
Such a result has been confirmed by \citet{Cap10} using the Swift-BAT all sky survey at z$\sim$0. They detected both a L$_X$
dependence of AGN clustering amplitude and a larger clustering of Type I AGN than that  of  Type II AGN. 
\citet{Kru10, Kru11} confirm the weak dependence of the clustering strength on AGN X-ray luminosity at a $2\sigma$ level for $z<0.5$.

\begin{figure}
\centering
\includegraphics[width=0.5\textwidth]{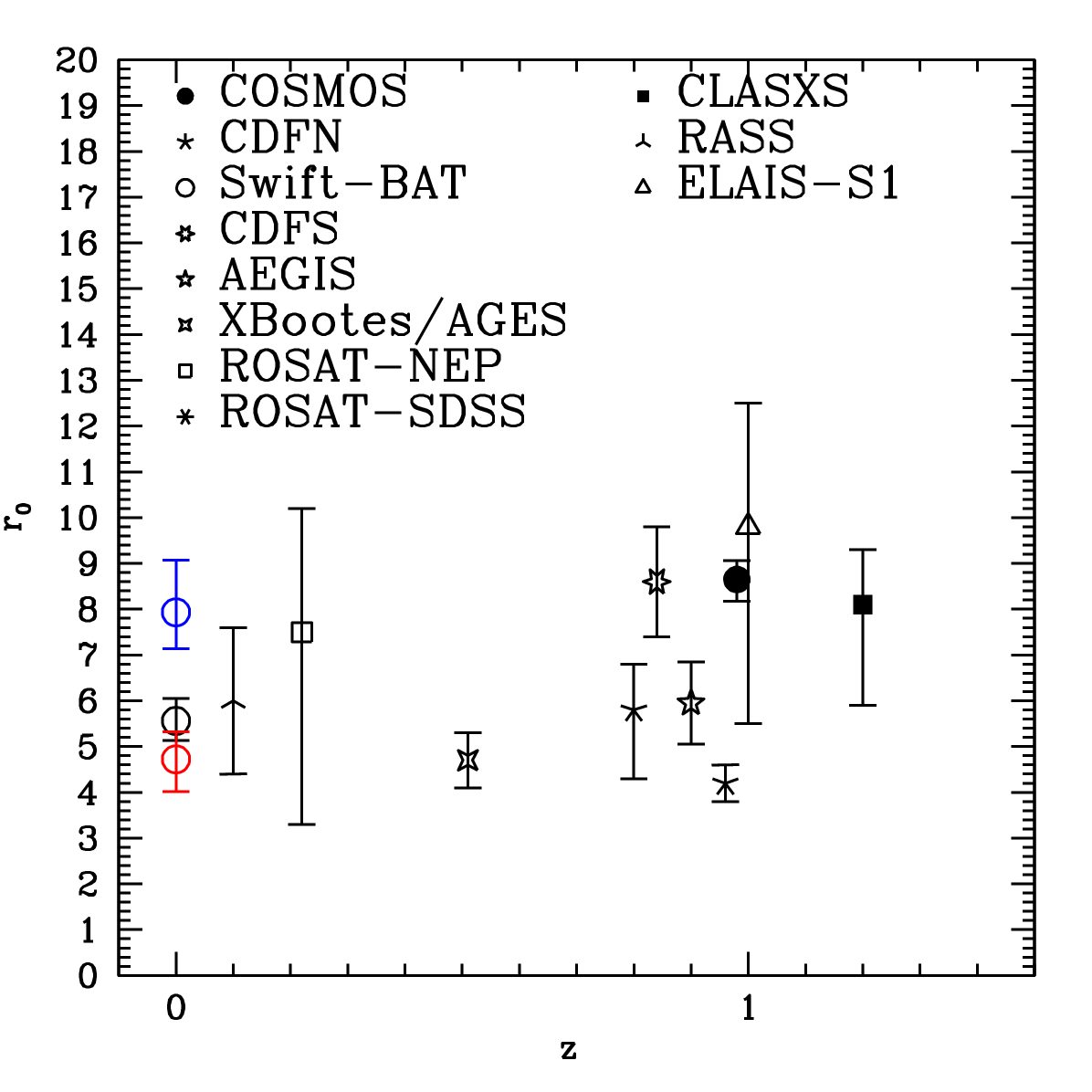}
\caption{\footnotesize Redshift evolution of the correlation length $r_0$ as estimated in different X-ray surveys 
(COSMOS, \citet{Gil09, Alle11}; CDFN, \citet{Gil05, Yan06}; Swift-BAT, \citet{Cap10}; CDFS, \citet{Gil05}; AEGIS, 
\citet{Coi09}; AGES, \citet{Hic09}; ROSAT-NEP, \citet{Mul04}; ROSAT-SDSS, 
\citet{Kru10}; CLASXS, \citet{Yan06}; RASS, \citet{a00}; ELAIS-S1, \citet{Puc06}).}
\label{fig:ro}
\end{figure}

Table \ref{tab1} summarizes all the discussed results on the clustering of AGN in X-ray surveys with  
bias factors converted to a common cosmology ($\Omega_{\Lambda}$ = 0.7,$\Omega_m$=0.3, $\sigma_8$=0.8) in the 
EMSS, \citet{b93} ; RASS, \citet{v95,a00}; ROSAT-NEP, \citet{Mul04}; AXIS, \citet{Ebr09}; ELAIS-S1, \citet{Puc06}; CDFS, \citet{Gil05}; CDFN, \citet{Gil05, Yan06}; XMM-2dF, \citet{Bas05}; XMM-LSS, \citet{Gan06}; 
CLASXS, \citet{Yan06}; COSMOS, \citet{Gil09, Alle11};  Swift-BAT, \citet{Cap10};  AEGIS, \citet{Coi09}; AGES, \citet{Hic09}; ROSAT-SDSS, \citet{Kru10},
while fig. \ref{fig:ro} shows the redshift evolution of the correlation length $r_0$ as estimated in previous works,
according to the legend.

\subsection{Techniques of investigation} 
The continuously increasing volume and quality of data, 
allowed a parallel improvement of the techniques of investigation. 
The first surveys  of  {\em Einstein}   \citep[see e.g.][]{xbf}, used the autocorrelation function 
of the unresolved CXB and linked it to the clustering properties of the clustering
of X-ray source that produced it. 

Modern surveys have mostly estimated correlation function with estimators that use
of random samples and real data pairs and then estimating  physical clustering 
properties by fitting the correlation function functions with simple power-law models
in the form of eq. \ref{pl}. A detailed description of the method to estimate 
correlation functions is given in the appendix.
Considering its power,  here we give a detailed description of halo modeling
which is by far the most reliable 
formalism to describe clustering of AGN/Galaxies and to 
determine the environment of a specific DMH  tracer.

\section{Halo Model}
\label{sec1}

In the hierarchical model of cosmological structure formation, 
galaxies, group of galaxies, clusters an so on are built from the 
same initial perturbation in the underlying dark matter density field. 
Regions of dark matter denser-than-average collapse to form halos 
in which structures form. 
Galaxies and AGN, as well as, groups and clusters are believed 
to populate the collapsed DMHs.

The theoretical understanding of galaxy clustering has been greatly 
enhanced through the framework of the \emph{halo model} 
\citep{Kau97,Pea00,Coo02,Tin05,Zhe05}.
One can fill DMHs with objects 
based on a statistical \emph{halo occupation distribution} (HOD), 
allowing one to model the clustering of galaxies within halos 
(and thus at non-linear scales) while providing a self-consistent determination 
of the bias at linear scales. 
Similarly the problem of discussing the abundance 
and spatial distribution of AGN can be reduced to studying 
how they populate their host halos. \\
The HOD analysis recasts AGN clustering measurements 
into a form that is more physically informative and conducive 
for testing galaxy/AGN formation theories.

Thus, one can use measurements of AGN two-point 
correlation functions to constrain the HOD of different sets of AGN and gain 
information on the nature  of DMH in which they live.
In fact, the power of the HOD modeling is the capability to transform 
data on AGN pair counts at small-scales into a physical relation between AGN 
and DMH at the level of individual halos.

The key ingredient needed to describe the clustering 
properties of AGN is their \emph{halo occupation distribution function}
$P_N(M_h)$, which gives the probability of finding $N$ AGN within 
a single halo as a function of the halo mass, $M_h$.
In the most general case, $P_N(M_h)$ is entirely specified by all 
its moments which, in principle, could be observationally determined 
by studying AGN clustering at any order. Regrettably AGN are so rare 
that their two-point function is already  poorly determined, so 
that it is not possible to accurately measure higher-order statistics. 
One overcomes this problem by assuming a predefined functional 
form for the lowest-order moments of $P_N(M_h)$, defining
the \emph{halo occupation number} $N(M_h)$ which is the mean value 
of the halo occupation distribution 
$N(M_h) = <N>(M_h) = \sum_N N P_N(M_h)$.
It is convenient to describe $N(M_h)$ in 
terms of a few parameters whose values will then be constrained by the data.

An accurate description of matter clustering on the basis of the
halo approach requires three major ingredients:
the halo mass function $n(M_h)$ (the number of DMHs
per unit mass and volume), the mass-dependent biasing
factor $b(M_h)$ and the density profile of halos.
These terms, along with a parametrization of $N(M_h)$, allow us to calculate 
some useful quantities; the number density of AGN:
\begin{equation}
n_{AGN} = \int n(M_h) N(M_h) dM_h
\end{equation}
the large-scale bias:
\begin{equation}\label{bHOD}
b = \frac{\int b_h(M_h) N(M_h)n(M_h)dM_h}{\int N(M_h)n(M_h)dM_h}
\end{equation}
and the average mass of the host dark halo:
\begin{equation}\label{MHOD}
M =	\frac{\int M_h N(M_h)n(M_h)dM_h}{\int N(M_h)n(M_h)dM_h}
\end{equation}

The number density and clustering properties of the DMHs can be 
easily computed, at any redshift, by means of a set of analytical tools 
which have been tested and calibrated against numerical simulations 
\citep{Mo96, She99, She01, Tin05, Bas08, Tin10,Pil10,ma11}.
Popular choices for both $n(M_h)$ and $b(M_h)$ 
are the analytical spherical collapse \citep{She99} or 
an ellipsoidal collapse model \citep[see \S\ref{sec4} for more details]{She01}.
A detailed description of HOD mathematical formalism is given in Appendix B.

\subsection{Occupation Number}

In the past ten years, a very successful framework for modelling 
the nonlinear clustering properties of galaxies has been developed and
a number of halo models have been presented in the literature. These 
have been successfully used to describe the abundance and clustering 
properties of galaxies at both low \citep{Pea00,Sel00,Sco01,Ber02, Mar02,Mag03, 
van03,Yan03, Zeh04, Tin05, Phl06,Zhe09} 
and high \citep{Bul02, Mou02, Ham04, Zhe04, Zhe07}
redshifts, as well as whether these galaxies occupy the centers of the DMH or are 
satellite galaxies \citep{Kra04,Zhe05}.
 
Partially due to the low number density of AGN, there have been few 
results in the literature interpreting AGN correlation function using HOD modelling, 
where the small-scale clustering measurements are essential. 
\citet{Por04} studied the clustering of 2QZ QSO with the halo model 
to infer the mean number of optically selected quasars which are harboured 
by a virialized halo of given mass and the characteristic quasar lifetime.
\citet{Pad09} discussed qualitative HOD constraints 
on their LRG-optical QSO Cross-Correlation Function (CCF) and \citet{She10} modelled with the HOD the observed 
two-point correlation function of 15 binary quasar at $z>2.9$.

The standard halo approach used for quasars and galaxies is based on the 
idea that the elements of HOD can be effectively 
decomposed into two components, separately describing the properties of 
central and satellite galaxies within the DMH.
A simple parametric form used to describe the galaxy HOD
is to model the mean occupation number for central galaxies as a step function, 
i.e., $ \langle N_{cen} \rangle = 1$ for halos with mass $M \geq M_{min}$ and $\langle N_{cen} \rangle = 0$
for $M < M_{min}$, while the distribution of satellite objects can be well 
approximated by a Poisson distribution with the mean following a power law, 
$\langle N_{sat} \rangle = (M/M_1 )^{\alpha}$. 
Previously derived HOD of galaxies show $\alpha$ values $\sim 1-1.2$ which
implies a number of satellite galaxies approximately proportional to $M_h$.

The clustering properties of X-ray selected AGN have been modelled with the HOD
in two previous works for sources in the \emph{Bootes} field \citet{Sta10} and 
in the \emph{ROSAT All-Sky Survey} \citet{Miy11}. 
\citet{Sta10} used the the projections of the two-points correlation function
both on the sky plane and in the line of sight to show that \emph{Chandra/Bootes} AGN are located at 
the center of DM halos with $M>M_{min}=4 \times 10^{12}$ h$^{-1}$ $M_{\odot}$,
assuming a halo occupation described by a step function (zero AGN per halo/subhalo
below $M_{min}$ and one above it). They also showed that Chandra/Bo\"{o}tes AGNs are located at the
centers of DMHs, limiting the fraction of AGN in non-central galaxies 
to be $< 0.09$ at the 95\% CL. The central locations of the AGN host galaxies 
are expected in the merger trigger model because mergers of equally-sized galaxies 
preferentially occur at the centers of DMH \citep{Hop08}.

\citet{Miy11} modelled the AGN HOD testing the effects of having or not
AGN in central galaxies by using the RASS AGN-LRG cross-correlation. 
In the first scenario they assumed that all the AGN are satellites and they
visualized the HOD of the LRG as a step function with a step at $log M_{h}[h^{-1}M_{\odot}=13.5]$.
While formally they assumed that all AGN are not in central galaxies,
the HOD constraints obtained from this assumption can be applied to
satellite and central AGN if the AGN activity in central galaxies of high-mass halos 
($log M_{h}[h^{-1}M_{\odot}>13.5]$) is suppressed.
In particular, they used a truncated power-law satellite HOD, with two
parameters: the critical DMH mass below which the AGN HOD is zero
and the slope $\alpha$ of the HOD for $M_h> M_{cr}$. They also
investigated a model where the central HOD is constant and the satellite HOD
has a power-law form, both at masses above $M_{min}$.
In all the cases they rejected  $\alpha \sim 1$, finding a marginal
preference for an AGN fraction among satellite galaxies which decreases
with increasing $M_h$. They argued that this result might be explained
by a decrease of the cross-section for galaxy merging in the 
environment of richer groups or clusters. In fact previous observations 
infer that the AGN fraction is smaller in clusters than in groups
\citep{Arn09, Mar09, Sil09,kou10}.

It is important to stress that the small number statistics has so far limited 
the accuracy of correlation function of X-ray AGN at small-scales, 
especially through the auto-correlation 
function of the AGN themselves. The situation can be improved by 
measuring the cross-correlation function of AGN with a galaxy sample 
that has a much higher space density, with common sky and redshift 
coverage as the AGN redshift surveys. The AGN clustering through 
cross-correlation function with galaxies is emerging in the last years
\citep{Li06, Coi07, Coi09, Hic09, Pad09, Mou09}
and can be used
to improve our understanding of how AGN populate DMH \citep{Miy11, Kru11}.

\section{Bias and DMH Mass}\label{sec4}

\begin{figure}
\centering
\plotone{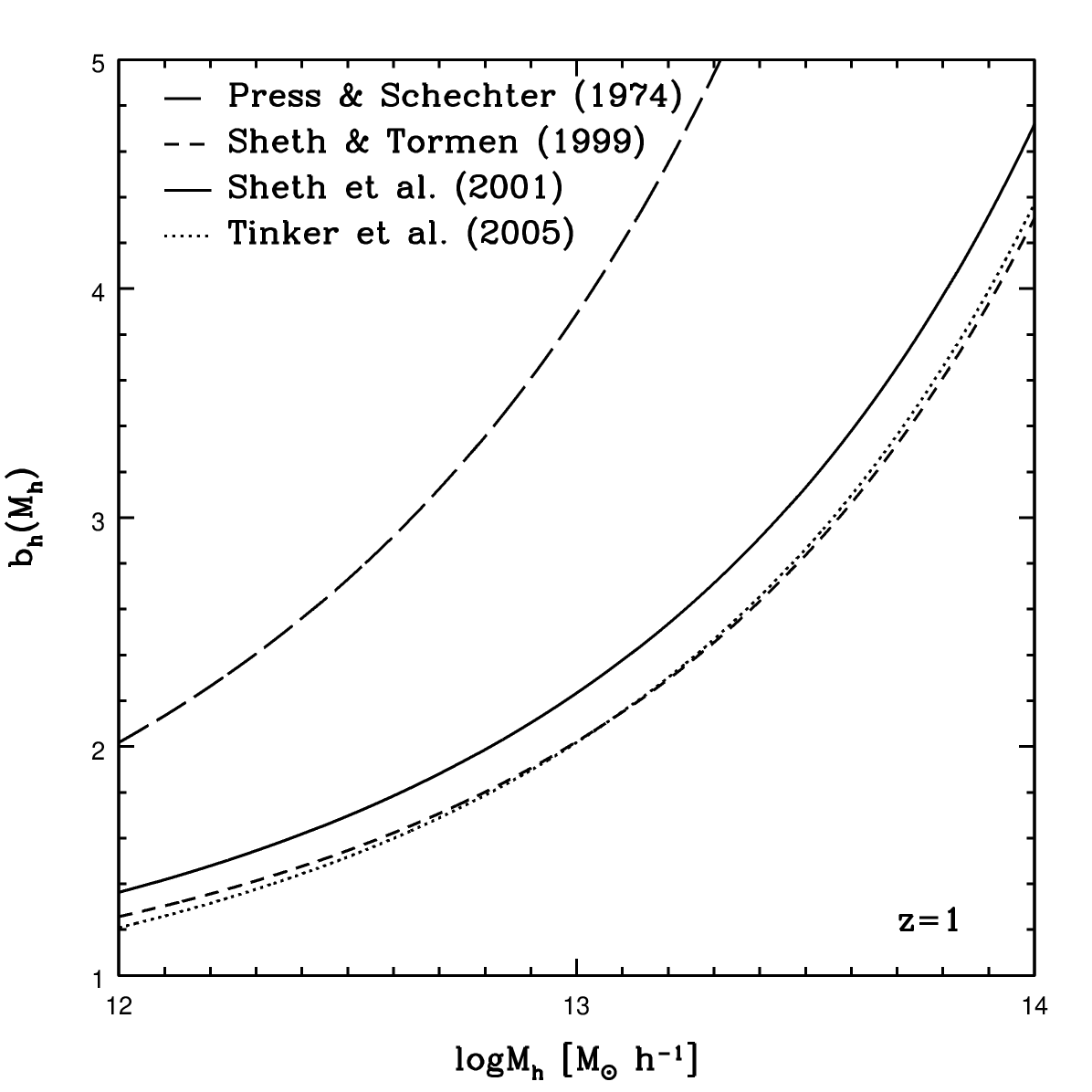}
\caption{\footnotesize Halo bias as function of halo masses for a fixed redshift 
              z=1 and the corresponding predictions of Press \& Shechter (1974) (\emph{long-dashed line}), 
              \citet [\emph{dashed line}]{She99} , \citet [\emph{solid line}] {She01} and \citet[\emph{dotted line}]{Tin05}.}
\label{fig:biasvsM}
\end{figure}

In the literature, the bias parameter is often calculated with the power-law
fits \citep{Mul04, Yan06, Miy07, Coi09, Kru10, Hic11} over scales
of 0.1-0.3 $<r_p<$ 10-20 $h^{-1}$ Mpc. The power-law
models of the ACF are usually converted to the rms fluctuation over 8 $h^{-1}$ Mpc
spheres or are averaged up to the distance of 20 $h^{-1}$ Mpc. 
While some authors use only large scales ($r_p>1-2$ $h^{-1}$ Mpc) to ensure that 
the linear regime is used, others include smaller scales to have better statistics.
As an example, \citet{Hic09} fitted their data with a biased DMH projected correlation function.\\
In the HOD analysis the bias factor only comes from the 2-halo term ($r_p>$1-2 $h^{-1}$ Mpc).
\citet{Miy11} compared the bias of RASS-AGN from the full HOD model (Eq. \ref{bHOD}) 
with the one estimated using the power-law best fits parameters, finding that
the bias estimates are consistent within 1$\sigma$. Moreover, 
using Eq. \ref{eq:PLb} one introduces large statistical errors.
\citet{Alle11} found a similar results in comparing the bias of X-ray AGN in COSMOS field
from the 2-halo term with Eq. \ref{eq:b} and the one estimated from the 
power-law best fits parameters.
In Appendix C, we describe the mathematical procedures for the bias parameter 
calculation commonly used in the literature.

Most of the authors \citep{Hic09, Kru10, Cap10}
used  an analytical expression \citep[as the one described in][]{She99,
She01, Tin05,Bas08} to assign a characteristic DMH mass to the hosting halos. 
The large-scale bias is directly related to the mass function of halos,
so that the mass of a halo dictates the halo clustering and the number
of such halos.
The halo mass can be quantified in terms of the peak height
$\nu = \delta_c/\sigma(M_h,z)$, which characterizes the amplitude
of density fluctuations from which a halo of mass $M_h$ form at a 
given redshift. In general one assumes $\delta_c=1.686$ and $\sigma(M_h,z)$
is the linear overdensity variance in spheres enclosing a mean mass $M_h$.
The traditional choice of the mass function and then of the bias
has been that of \citet{Pre74}:
\begin{equation}\label{eq:biasPS}
b^{PS}=1+\frac{\nu^2-1}{\delta_c}
\end{equation}
A commonly-used prescription was derived by \citet{She99}:
\begin{equation}\label{eq:biasST}
b^{ST}=1+ \frac{a\nu^2-1}{\delta_c} + \frac{2p/\delta_c}{1+(a\nu^2)^p}
\end{equation}
where $a=0.707$ and $p=0.3$
or the ellipsoidal collapse formula of \citet{She01}:
\begin{eqnarray}\label{eq:biasMTS}
b^{SMT}=1+\frac{1}{\sqrt{a}\delta_c} [ \sqrt{a}(a\nu^2) +   \nonumber \\
\sqrt{a}b(a\nu^2)^{1-c} - \frac{(a\nu^2)^{c}}{(a\nu^2)^{c}}   \nonumber \\
 + b(1-c)(1-c/2) ] 
\end{eqnarray}
where $a=0.707$, $b=0.5$, $c=0.6$ or the recalibrated parameters
$a=0.707$, $b=0.35$, $c=0.8$ of \citet{Tin05}.
The $\nu$ parameter can be estimated following the Appendix of \citet{van02}. 
Fig. \ref{fig:biasvsM} shows the bias as function of the halo mass $M_h$, at z=1,
following the predictions of \citet{Pre74, She99, She01, Tin05}.

\citet{Alle11} argued that this approach reveals an incongruity due
to the fact that the AGN bias used in the formulas above, is the
average bias of a given AGN sample at a given redshift. 
In fact,  following this approach one can not take into account that the 
the average bias is sensitive to the entirety of the mass distribution;
different mass distributions with different average masses can give rise to 
the same average bias.

On the contrary by using the halo model, the average bias and the average mass of the sample,
 Eq. \ref{bHOD} and Eq. \ref{MHOD} properly account for the shape of the mass distribution:
the average bias depends on the halo number density and on the AGN HOD, 
integrated over the mass range of the particular AGN sample.
They introduced a new method that uses the 2-halo
term in estimating the AGN bias factor assuming an AGN HOD described by
a $\delta$-function. Following this approach they  
properly took into account for the sample variance and the growth
of the structures over time associated with the use of large redshift interval
of the AGN sample.

On the other hand, \citet{Miy11} and \citet{Kru11} applied the HOD modeling technique to the RASS AGN-LRG CCF in order to 
move beyond determining the typical DMH mass based on the clustering signal strength and instead 
constrain the full distribution of AGN as a function of DMH mass. 
Along with a parametrization of $N(M_h)$ they estimated the large-scale bias and the typical mass 
of hosting DM halos using Eq. \ref{bHOD} and Eq. \ref{MHOD}. This method improves the clustering analysis 
because it properly uses the non-linear growth of matter in the \textit{1-halo} term through the formation and growth of DMHs. 
These results are significant improvements with respect to  the standard method of fitting 
the signal with a phenomenological power law or using the 2-halo term (see Appendix C).
 
\begin{figure*}
\plottwo{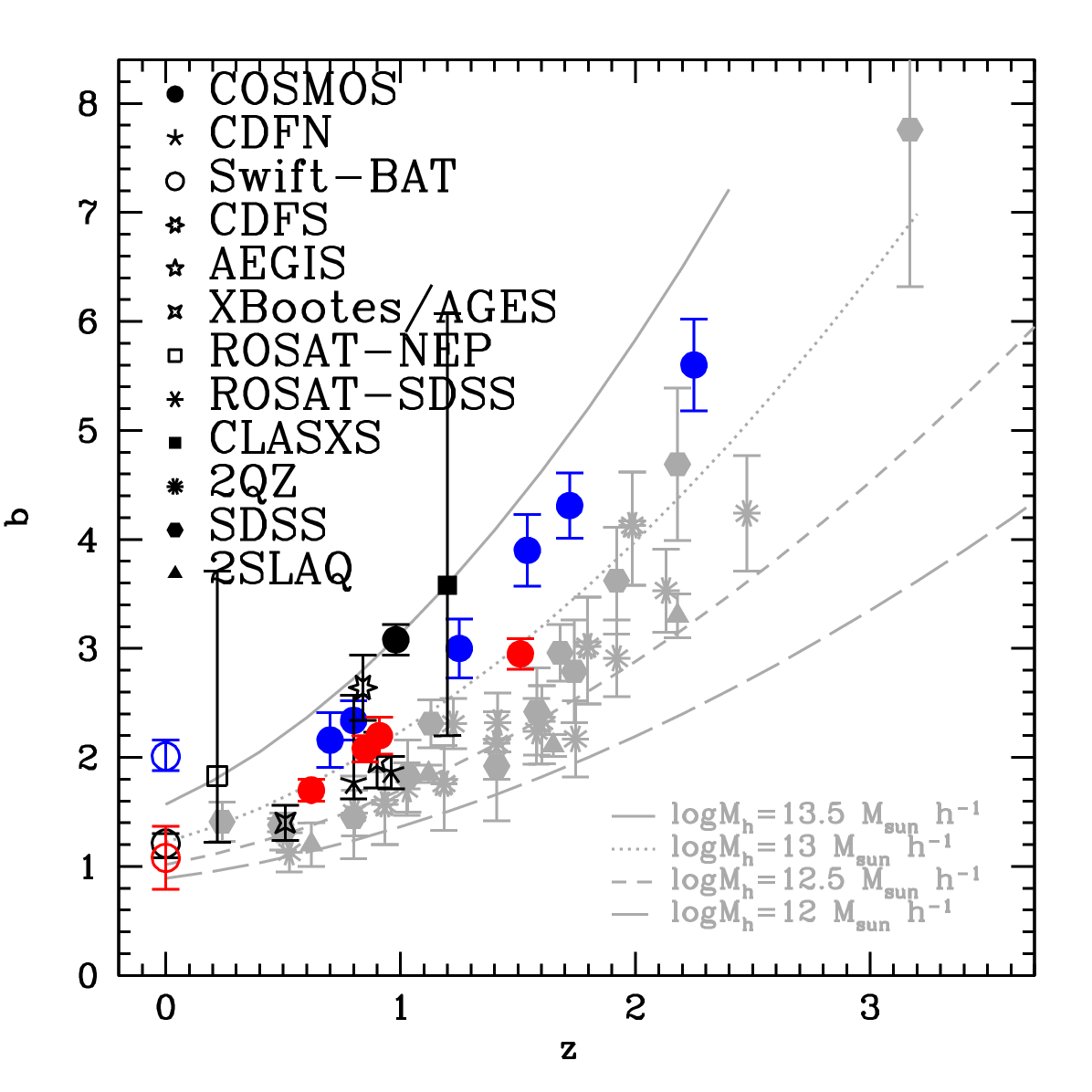}{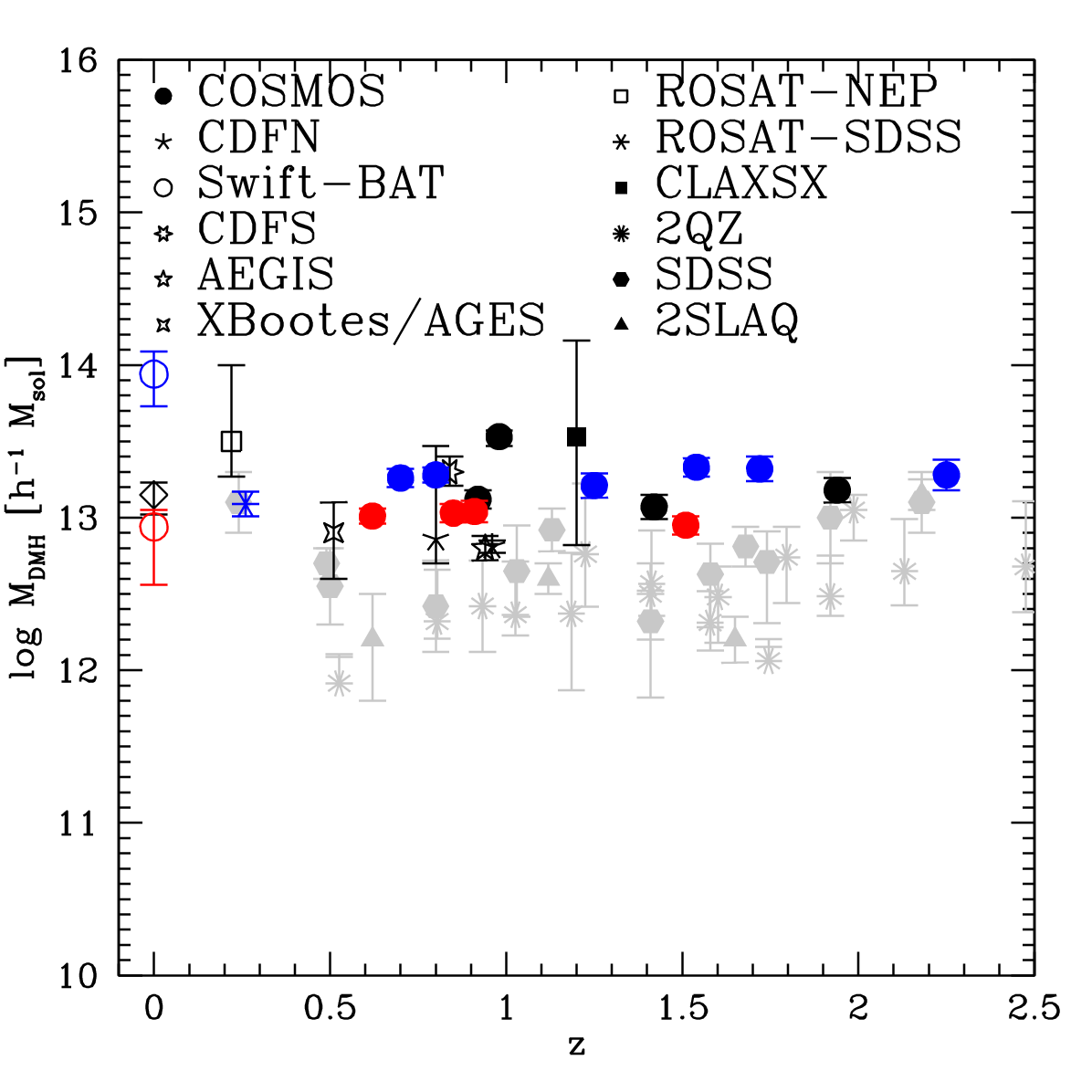}
\caption{\footnotesize Bias factor (\emph{Left Panel}) and mass of AGN hosting halos
 (\emph{Right Panel}) as a function of redshift for
  X-ray selected AGN (black data points), X-ray selected Type 1 AGN
  (blue data points) and X-ray selected Type 2 AGN (red data points) as estimated in different surveys
  (COSMOS, \citet{Gil09, Alle11}; CDFN, \citet{Gil05, Yan06}; Swift-BAT, \citet{Cap10}; CDFS, \citet{Gil05}; 
  AEGIS, \citet{Coi09}; AGES, \citet{Hic09}; ROSAT-NEP, \citet{Mul04}; ROSAT-SDSS, \citet{Kru10}; CLASXS, \citet{Yan06}). 
  The dashed lines show the expected
  $b(z)$ of DMHs with different masses according to the legend, based on Sheth et al. (2001).
   The grey points show results from quasar - quasar correlation measurements using spectroscopic 
   samples from SDSS \citep{Ros09,Shen09}, 2QZ \citep{Cro05,Por06} and 2SLAQ \citep{Ang08}.
All the previous studies infer the picture that X-ray selected AGN which are
moderate luminosity AGN compared to bright quasars,
inhabit more massive DMHs than optically selected quasars in the range $z=0.5-2.25$.}
\label{fig:biasvsz}
\end{figure*}

\subsection{ X-ray selected AGN bias, bias evolution and mass of the  hosting halos}
\label{sec2.2}

The majority of the X-ray surveys agree with a picture where X-ray AGN
are typically hosted in DM halos with mass of the order
of 12.5$<logM_{DMH}$ $[h^{-1}M_{\odot}]<$13.5, at low (z$<$0.4) 
and high (z$\sim$1) redshift \citep{Gil05, Yan06, Gil09, Hic09, Coi09, Kru10, Cap10, Sta10, Miy11, Kru11}.

At high redshift, \citet{Gil05} 
measured the clustering of X-ray AGN with z=0-4 
in both the $\sim$ 0.1 $deg^2$ CDFs,
finding $b=1.87^{+0.14}_{-0.16}$ for 240 sources in the northern field and
$b=2.64^{+0.29}_{-0.30}$ for 124 sources in the southern field.
At $z \sim 1$, \citet{Yan06} 
measured the clustering of 233 spectroscopic sources in the
0.4 $deg^2$ \emph{Chandra} CLASXS area and of 252 spectroscopic sources from the CDFN,
both at z=0.1-3. They found $b=3.58^{+2.49}_{-1.38}$ for the CLASXS AGN and
$b=1.77^{+0.80}_{-0.15}$  for the CDFN field.\\
\citet{Gil09} 
studied 538 XMM-COSMOS AGN
with 0.1 $< z <$ 3 and they found a bias factor 
$b=3.08^{+0.14}_{-0.14}$ at $\overline{z} \sim 1$. Using the
Millennium simulations they suggested that XMM-COSMOS AGN 
reside in DMH with mass M$_{DMH} > 2.5 \times 10^{12} h^{-1}$ M$_{\odot}$.
\cite{Coi09} 
measured the clustering of X-ray AGN at z=0.7-1.4 in the AEGIS field
and they estimated $b=1.85^{+0.28}_{-0.28}$. Following \citet{Zhe07}
they infer from the bias factor that at $z=0.94$ the minimum DM
halo mass of the X-ray AGN is $> 10^{12} M_{\odot}h^{-1}$.  
These results combined with \citet{Mou11} show that moderate luminosity X-ray 
selected AGN live in DMHs with masses $M_h \sim 10^{13} h^{−1} M_{\odot}$
at all redshifts since $z \sim 1$.
At lower redshift 
\citet{Hic09} analysed 362 AGES X-ray AGN at
$<z>$=0.51. The bias factor equal to $b=1.40 \pm 0.16$ indicates
that X-ray AGN inhabit DM halos of typical mass $\sim 10^{13}
M_{\odot}h^{-1}$. 

In the local Universe \citet{Cap10} estimated for $\sim$ 200
Swift-BAT AGN a bias equal to $b=1.21^{+0.07}_{-0.06}$ which
corresponds to log$M_{DM}$=13.15$^{+0.09}_{-0.13}$ h$^{-1}$ $M_{\odot}$.\\
\citet{Alle11} estimated an average 
mass of the XMM-COSMOS AGN hosting halos equal to
$logM_{0}[h^{-1}Mpc]=13.10 \pm 0.06$ at $z \sim 1.2$.
They also measured the bias of Type 1 and Type 2 AGN, finding that the latter
reside in less massive halos than Type 1 AGN.
Only two other works \citep{Cap10,Kru10} analysed 
the clustering properties of X-ray selected Type 1 AGN and Type 2 AGN.
\citet{Cap10} estimated the typical DM halo
mass hosting type 1 and type 2 Swift-BAT AGN at $z \sim 0$. They measured that 
these two different samples are characterized
by halos with mass equal to $logM_{DM}[h^{-1}
M_{\odot}] \sim 13.94^{+0.15}_{-0.21}$ and $\sim 12.92^{+0.11}_{-0.38}$, respectively.
However the lack of small separation pair of Type I AGN in the local Universe 
may have produced systematic deviations which were not accounted in their fits. 
In \citet{Kru10} the bias factor of BL RASS AGN at
$z=0.27$ are consistent
with BL AGN residing in halos with mass $logM_{DM}[h^{-1} M_{\odot}] = 12.58^{+0.20}_{-0.33}$.\\
Using the HOD model, \citet{Sta10} suggested that X-ray 
Chandra/Bootes AGN are located at the center of DM halos with $M > M_{min} = 4
\times 10^{12} h^{-1} M_{\odot}$ while \cite{Miy11} estimated 
for RASS AGN at z=0.25 $b=1.32\pm 0.08$ and a typical mass of the host halos of
$13.09\pm 0.08$.

The redshift evolution of the clustering of X-ray selected AGN has been first studied
by \citet{Yan06} in the CLAXS+CDFN fields. They measured an increase of the bias factor
with redshift, from $b=0.95\pm0.15$ at z=0.45 to $b=3.03\pm0.83$ at z=2.07,
corresponding to an average halo mass of $\sim$12.11 h$^{-1}$ $M_{\odot}$.

\citet{Alle11} studied the redshift evolution of the bias for a sample
of XMM-COSMOS AGN at $z<2$. They found a bias evolution with time 
from $b(z=0.92)=1.80 \pm0.19$ to
$b(z=1.94)=2.63\pm0.21$ with a
DM halo mass consistent with being constant at $logM[h^{-1} M_{\odot}]\sim 13.1$
at all redshifts $z<2$.
They also found evidence of 
a redshift evolution of the bias factor
of XMM-COSMOS Type 1 AGN 
and Type 2.
The bias evolves with redshift at constant average halo mass $logM_{0}[h^{-1} M_{\odot}] \sim 13.3$ for 
Type 1 AGN  and $logM_{0}[h^{-1} M_{\odot}] \sim 13$ for Type 2 AGN 
at $z<2.25$ and $z<1.5$, respectively.
In particular \citet{Alle11} argued that X-ray selected Type 1 AGN reside in more massive
DMHs compared to X-ray selected Type 2 AGN at all redshifts at $\sim 2.5\sigma$ level,
suggesting that the AGN activity is a mass
triggered phenomenon and that different AGN classes
are associated with the DM halo mass, irrespective of redshift $z$. 

\citet{Kru11} measured the clustering amplitudes of both X-ray RASS and optically-selected SDSS 
broad-line AGNs, as well as for X-ray selected narrow-line RASS/SDSS AGNs through cross-correlation functions 
with SDSS galaxies and derive the bias by applying the HOD model directly to the CCFs.
They estimated typical DMH masses of broad-line AGNs in the range $log(M_h/[h^{−1} M_{\odot}])= 12.4-13.4$, 
consistent with the halo mass range of typical non-AGN galaxies at low redshifts and 
they found no significant difference between the clustering of X-ray selected narrow-line AGNs and broad-line AGNs
up to $z\sim0.5$.

Fig. \ref{fig:biasvsz} shows the bias parameter (\emph{Left Panel}) and the mass of the AGN hosting halos (\emph{Right Panel})
as a function of redshift for X-ray selected AGN (black data points), X-ray selected Type 1 AGN (blue data points) and 
X-ray selected Type 2 AGN (red data points) as estimated for different surveys (see the legend).
The dashed lines show the expected b(z) of typical DM halo masses $M_{DMH}$ based on Sheth et al. (2001). 
The masses are given in logM$_{DMH}$ in units of $h^{-1}$ M$_{\odot}$.

There have been several studies of the bias evolution of optical quasar
with the redshift as shown in fig. \ref{fig:biasvsz} (grey data points), 
based on large survey samples such as 2QZ, 2SLAQ and SDSS \citep{Cro05, Por06, Shen09, Ros09, Ang08}. 
These previous studies infer the picture that X-ray selected AGN which are
moderate luminosity AGN compared to bright quasars,
inhabit more massive DMHs than optically selected quasars in the range $z=0.5-2.25$.\\
Recently, \citet{Kru11} verified that the clustering properties 
between X-ray and optically- selected AGN samples are not significantly different
in three redshift bins below $z=0.5$ (the differences are $1.5\sigma$, 0.1$\sigma$ and 2.0$\sigma$).The reason 
for the fact that X-ray selected AGN samples appear to cluster 
more strongly than optically- selected AGNs is still unclear.
\citet{Alle11, Mou11} suggested that the difference in the bias and then in the host DMH masses 
is due to the different fueling mode of those sources from that of the X-ray selected moderate luminosity AGN.
On the contrary, \citet{Kru11} suggested that some of the X-ray clustering studies significantly underestimate 
their systematic uncertainties and then it may turn out that these 
measurements are consistent with optical AGN clustering measurements.
More high-z AGN clustering measurements based on larger samples are needed to gain a clearer picture.

\subsection{AGN Life Time} 

One of the most important tests for studying the  evolution 
models of AGN is understanding their lifetime.  
It is widely accepted that AGN is
phase of the galaxy life necessary to explain the coevolution
of the bulge and the black hole. After a triggering event, of which 
we do not know the nature, yet, the central black holes
begins its accretion phase and it is believed that undergoes several
regimes of Eddington rates and bolometric luminosity. 
\citet{party} proposed a method to derive the AGN life time 
by knowing their  space density and their DMH host mass. \\
By knowing the AGN and DMH halo space density at a given luminosity and mass (n$_{AGN}$, n$_{DMH}$),
 one can estimate the duty cycle of the AGN, $\tau _{\rm AGN}(z)=\frac{n_{\rm AGN}(L,z)}{n_{\rm DMH}(M,z)}\tau _H(z)$, 
 where $\tau(H(z))$ is the Hubble time at a given redshift.
 Actually this method provides only an upper limit since it assumes that the life
 of an halo of a given mass is similar to the Hubble time. 
 A more exhaustive formulation would $\tau _{\rm AGN}(z)=\frac{n_{\rm AGN}(L,z)}{n_{\rm DMH}(M,z)}\tau _{DMH}(z)$,
 where $\tau _{DMH}(z)$ is the age of a DMH at given redshift. Unfortunately this quantity 
 cannot be estimated analytically but could be estimate in a statistical way by using 
 hidrodynamic  simulations. 
 Several results can be mentioned for this quantities but their dispersion is very large, therefore we report
 only some example.
 At z=1, \citet{Gil09} obtains that the typical duty cycle of AGN is $<$1 Gyr. At z=0, \citet{Cap10} has measured 
 a duty cycle in the range 0.2 Gyr-5 Gyr with an expectation value of 0.7 Gyr. 
Both the measurements are fairly larger than the 40 million years  determined by \citet{party} at z=2-3. 
These differences however are not surprising if we assume that the different populations of AGN,
grow with a different Eddington rate as function of their typical luminosities and/or redshifts \citep{fab}.

\section{Discussion}

In this paper we reviewed the results in the field of X-ray AGN clustering, 
for energies between 0.1 keV to 55 keV over a period of more that 20 years. 
The literature has produced an increasingly convincing and consistent picture of
the physical quantities derivable from this kind of study. 
Most of the advancements in the field have been achieved with the 
improvement of survey capabilities and instruments sensitivity. 
The availability of simultaneously wide and deep fields, coupled with 
multi-wavelength information, has produced larger and larger samples
of spectroscopically confirmed sources. 
This allowed several teams to refine the techniques needed to estimate the 
two point ACF and the quantities derived form it. 
In particular we are entering a phase where, at least at z$<2$, AGN clustering studies 
won't probably provide any new result unless evaluated with the HOD formalism.
Open questions as what is the AGN occupation number  and the evolution of
HOD define a new barrier which is necessary
 to break in order to understand the history of X-ray emission 
from accretion onto AGN. In this respect, samples of X-ray selected AGN
always need a spectroscopical follow-up to provide a solid base to compute 
clustering in the real space rather than in the angular space. 

Summarizing,  the current picture is that X-ray selected AGN are highly biased
objects with respect to the underlined matter distribution. 
Such an evidence is clearer when measuring 
the redshift dependence  of AGN bias. 
At every redshift from z=0 to z=2, AGN cluster in way similar to DMH of 
mass of the order of log($M_{\odot}h^{-1}$)=13. The spread of such a value 
is of the order 0.3-0.5 dex at 1$\sigma$. 
This means that the determination of what kind of environment 
is inhabited by AGN, is relatively well constrained and identical 
at every redshift sampled by X-ray surveys.  This 
allows us to formulate the hypothesis that
every phase of  AGN activity is mass triggered phenomenon
(i.e. each AGN evolutionary phase is characterized by a critical halo mass).

It is believed that major mergers of galaxies is one of the dominant mechanisms for fueling quasars
at high redshift and bright luminosities, while
minor interactions, bar instabilities or 
tidal disruptions are important at low redshift ($z\lesssim 1$) and low luminosities
($L\lesssim 10^{44}erg$ s$^{-1}$) \citep{Hop06,Hop08a,Has08,Hop09}.
In the local Universe, for example, the study of the environment of Swift BAT Seyfert galaxies 
\citep{Kos10} finds a larger fraction of BAT AGNs
 with disturbed morphologies or in close physical pairs ($<$30 kpc)
  compared to matched control galaxies or optically selected AGNs. 
  The high rate of apparent mergers (25\%) suggests that AGN activity
   and merging are critically linked for the moderate luminosity AGN in the BAT sample. 
 Moreover models of major mergers appear
to naturally produce many observed properties
of quasars, as the quasar luminosity density, the shape and the evolution
of the quasar luminosity function and the large-scale quasar clustering as
a function of $L$ and $z$ \citep[e.g.,][]{Hop08, Shen09a, Sha09, Sha10, Sha10rev, Bon09, Tre10}. 
Quasar clustering at all redshift is consistent with halo masses similar to group scales,
where the combination of low velocity dispersion and moderate galaxy space density 
yields to the highest probability of a close encounter \citep{Hop08,apple}.  
Moreover recent detections of an L$_{X}$ dependent clustering
play in favor of major mergers being the dominant AGN triggering mechanism.

On the other hand it has became clear that many AGN are not fueled by major mergers and only
a small fraction of AGN are associated with morphologically disturbed
galaxies.
\citet{Geo07} and \citet{Sil09} found that AGN
span a broad range of environments, from the field to massive groups and thus 
major mergers of galaxies, possibly relevant for the more luminous quasar phenomenon, 
may not be the primary mechanism for fueling these moderate luminosity AGN.

\citet{Geo09} suggest that bar
instabilities and minor interactions are more efficient in producing
luminous AGN at $z\lesssim 1$ and not only Seyfert galaxies and
low-luminosity AGN as the \citet{Hop09} model predicts.
\citet{Cis10} analysed  a sample of X-ray selected AGN host galaxies
and a matched control sample of inactive galaxies in the COSMOS field.
They found that mergers and interactions involving AGN hosts are not dominant 
and occur no more frequently than for inactive galaxies. Over 55\% of the studied
AGN sample that is characterized by $L_{BOL} \sim 10^{45} erg$ $s^{-1}$ and by 
mass of the host galaxies $M_{\ast} \gtrsim 10^{10}M_{\odot}$
are hosted by disk-dominated galaxies, suggesting that secular fuelling 
mechanisms can be highly efficient. \\ 
Moreover several works on the AGN host galaxies \citep{Dun03, Gro05,
Pie07, Gab09, Rei09, Tal09} show that the morphologies of the AGN 
host galaxies do not present a preference for merging systems. 

At high redshift ($z \sim$ 2) recent findings of \citet{Sch11} and 
\citet{Ros11}, who examined a smaller sample of AGN
in the ERS-II region of the GOODS-South field,
inferred that late-type morphologies are prevalent among the AGN hosts.
The role that major galaxy mergers play in triggering AGN activity at 1.5 $< z <$ 2.5
was also studied in the CDF-S. 
At z=1.5-3 \citet{Scha11} showed that for X-ray-selected AGN in the Chandra Deep Field South and 
with typical luminosities of 10$^{42}$ erg s$^{-1}<$L$_X<$10$^{44}$ erg s$^{-1}$  the majority (~80\%) of the host galaxies of these 
AGNs have low SŽrsic indices indicative of disk-dominated light profiles, suggesting that secular processes govern 
a significant fraction of the cosmic growth of black holes. That is, many black holes in the present-day 
universe grew much of their mass in disk-dominated galaxies and not in early-type galaxies or major mergers. 

Later, \citet{Koc11} found that
X-ray selected AGN at $z\sim 2$ do not exhibit a significant excess 
of distorted morphologies while a large fraction reside in late-type galaxies.
They also suggested that these late-type galaxies are fueled by the stochastic 
accretion of cold gas, possibly triggered by a disk instability or minor interaction.

\citet{Alle11} argued that for moderate luminosity X-ray AGN 
 secular processes such as tidal disruptions or disk instabilities 
 might play a much larger role than major mergers up to $z \sim 2.2$.\\
It becomes important to study the clustering properties of AGN at high
redshift when we assume the peak of the merger-driven accretion. Moreover
given the complexity of AGN triggering, a proper selection of AGN samples, 
according to the luminosity or the mass of the host galaxies
can help to test a particular model boosting the fraction of AGN host 
galaxies associated with morphologically disturbed galaxies.

From the evolutionary point of view the evidence of 
a bias segregation of optically and X-ray selected AGN might be a sufficient
proof to claim that the two phenomena are sensitive 
to different environments and therefore
likely driven by different triggering mechanisms.
A more comprehensive picture will be available when the clustering 
of different phases of AGN activity will be studied and compared. 

\begin{figure*}
\centering
\includegraphics[angle=270,width=0.9\textwidth]{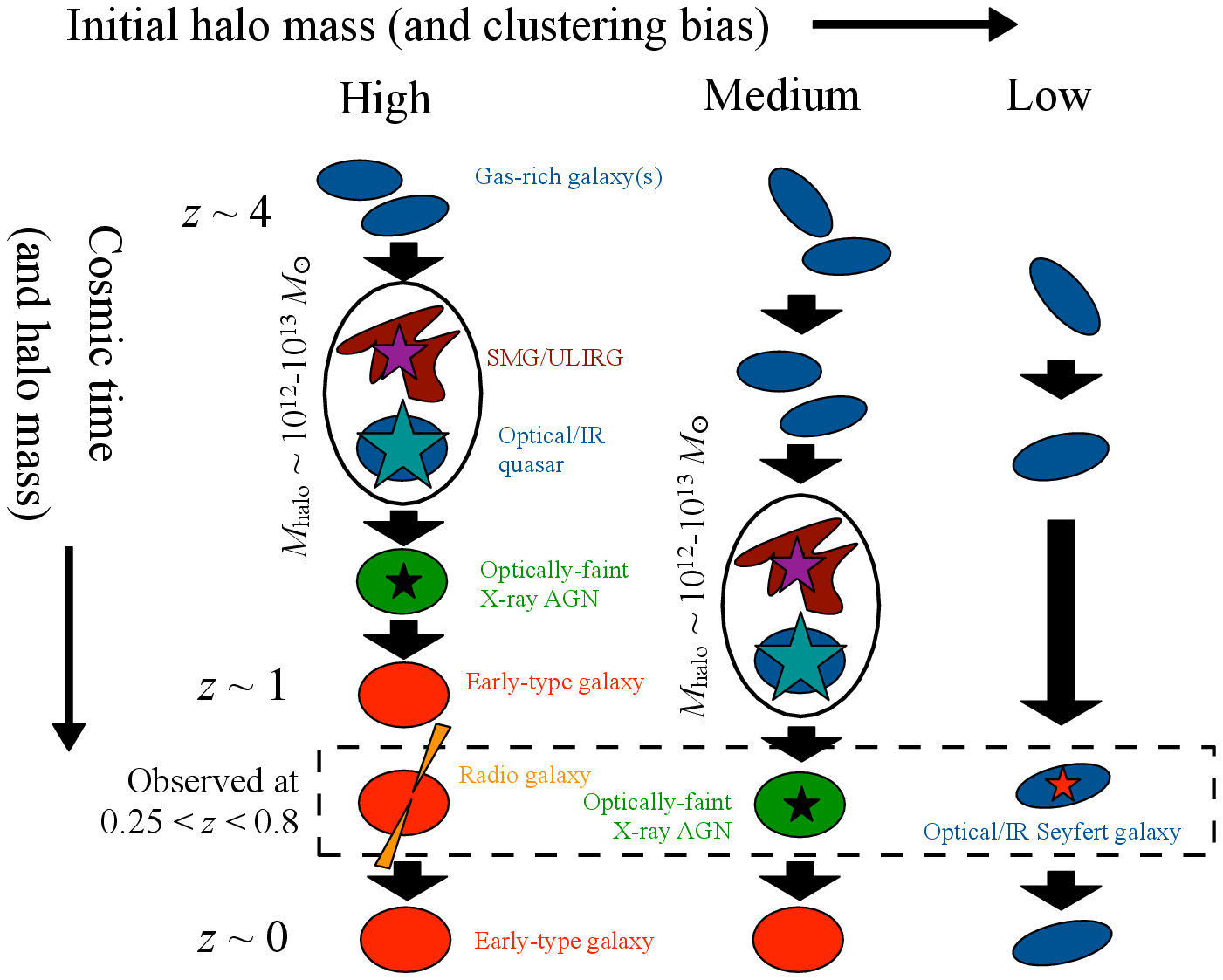}
\caption{\footnotesize Schematic for a simple picture of AGN and host galaxy
evolution, taken from \citet{Hic09}, and motivated by the
AGN host galaxy and clustering results presented in that study.}
\label{fig:hic}
\end{figure*}

\citet{Hic09} interpreted their clustering results in terms
of a general picture for AGN and galaxy evolution which is  reproduced in Fig. \ref{fig:hic}.
The picture consists of an evolutionary sequence that
occurs at different redshifts for halos with different masses.  In
this scenario, luminous AGN accretion occurs preferentially (through a
merger or some secular process) when a host DMH reaches a
critical $M_{DMH}$ between $10^{12}$ and $10^{13}$ M$_{\odot}$ h$^{-1}$ (this
phase is indicated by the solid ovals).  Once a large halo reaches
this critical mass, it becomes visible as a ULIRG or SMG (owing to a
burst of dusty star formation) or (perhaps subsequently) as a
luminous, unobscured quasar.  
The ULIRG/quasar phase is associated
with rapid growth of the SMBH and formation of a stellar spheroid, and
is followed by the rapid quenching of star formation in the galaxy.
Subsequently, the young stellar population in the galaxy ages
(producing "green" host galaxy), and the galaxy experiences
declining nuclear accretion that may be associated with an X-ray AGN.
Eventually the aging of the young stars leaves a "red" and "dead"
early-type galaxy, which experiences intermittent "radio-mode" AGN
outbursts that heat the surrounding medium.  For "medium" initial
DMHs, the quasar phase and formation of the spheroid
occurs later than for the systems with high halo mass, so that at
$z\sim0.5$ we may observe the green X-ray AGN phase. Even smaller
halos never reach the threshold mass for quasar triggering; these
still contain star-forming disk galaxies at $z\lesssim 0.8$, and we
observe some of them as optical or IR-selected Seyfert galaxies.  The
dashed box indicates the AGN types (in their characteristic DMH) 
that would be observable in the redshift range $0.25<z<0.8$. 

Further steps in the field will require the study of clustering of AGN from z=3 to z=6-7. 
This will likely lead to the determination of the mass of early DM spheroids who 
hosted primordial black holes seeds. However this is a very challenging task 
since it requires a very deep and wide survey with an almost complete optical follow-up. 

BOSS \citep{Eis11} and BigBOSS \citep{Sch11} will detect high 
redshift AGNs at $z \sim 2.2$, which will improve AGN clustering measurements at higher redshifts.
The only approved  mission that at the moment will allow to study the z=3-5 X-ray 
Universe is eROSITA \citep[launch Dec. 2013]{Pre07} for which an estimate of the completeness
of the typical follow up is still unavailable. Additionally, the Large Synoptic Survey Telescope \citep[LSST]{Ive08}
is expected to identify $\sim$2 million AGNs in optical bands. eROSITA and LSST have the potential to significantly 
improve AGN clustering measurements at low and high redshifts, though only if there are dedicated large 
spectroscopic follow-up programs. Another strong contribution will come form either Nustar that will likely 
provide a better view of AGN clustering without the selection biases introduced by photoelectric absorption. 
Athena the proposed ESA new  generation telescope that will mount a wide field imager on a very large collecting 
area telescope, will provide a further view on the deep X-ray sky and likely push our knowledge of the high-$z$ X-ray
Universe. 

In addition to better model the evolution of 
SMBH environments a fundamental point 
to start is to establish the nature of BH seeds at z=10.  Such a determination will likely
come with the new generation of telescope like JWST and ESO-ELT. 

\acknowledgements
NC thanks the INAF-Fellowship program for support. NC thanks the Della Riccia and 
Blanceflor-Lodovisi-Boncompagni foundation for partial support. 
VA is supported by the DFG cluster of excellence âOrigin and Structure of the Universeâ (www.universe-cluster.de).
NC, VA and AF thank the referees, Ryan Hickox and Manolis Plionis, for valuable
suggestions for improving the paper.
\clearpage
\appendix

\section{ Deriving the two points auto-correlation function}

The two-point auto-correlation function ($\xi(r)$, ACF)
describes the excess probability over random of finding a pair 
with an object in the volume $dV_1$ and another in the volume $dV_2$, separated by a distance $r$ so that
$dP=n^2[1+\xi(r)]dV_1 dV_2$, where $n$ is the mean space density. A known effect when measuring pairs separations is 
that the peculiar velocities combined with the Hubble flow
may cause a biased estimate of the distance
when using the spectroscopic redshift.  To avoid this effect
it is usually computed the projected ACF \citep{Dav83}:  
 $w(r_p)=2\int_{0}^{\pi_{max}}\xi(r_p,\pi)d\pi$.
Where $r_p$ is the distance component perpendicular to the line of sight and 
$\pi$ parallel to the line of sight (Fisher et al. 1994).
 It can be demonstrated that, if the ACF is expressed as 
$\xi(r)=(r/r_0)^{-\gamma}$, then
\begin{equation}
  w(r_{\rm p})=A(\gamma)r_{0}^{\gamma}r_{p}^{1-\gamma},
\label{eq:chi}
\end{equation}
 where  $A(\gamma)=\Gamma(1/2)\Gamma[(\gamma-1)/2]/\Gamma(\gamma/2)$ (Peebles 1980).\\
The ACF is mostly estimated by using the minimum variance estimator 
described by Landy \& Szalay (1993):
\begin{equation}
 \xi(r_p,\pi)=\frac{DD-2DR+RR}{RR}
\label{eq:landy}
\end{equation}
where DD, DR and RR are the normalized  number of
data-data, data-random, and random-random source pairs, respectively. 
Equation \ref{eq:landy}  indicates that an accurate estimate of the 
distribution function of the random  samples is crucial in order to 
obtain a reliable  estimate of $\xi(r_p,\pi)$. 
Note that other estimator have been proposed in the literature, but
the  \citet{Lan93} one has been shown to provide the smallest statistical variance. 
Such a formalism can be easily adopted when computing the angular or the redshift 
space correlation function, with the only difference that the evaluation is made on  a single 
dimension. 
Several observational biases must be taken 
 into account when generating a random sample of objects in a X-ray flux limited survey. 
In particular, in order to reproduce the selection function of the survey,
one has to carefully reproduce the space and  flux distributions 
of the sources, since the sensitivity in X-ray surveys
 is not homogeneous on the detector and therefore on the sky.
This points out the necessity of create a random sample which includes as many selection effects
as possible since the estimate of $\xi(r)$ (or $w(\theta)$) is strongly dependent on RR (see eq. 
\ref{eq:landy}). Moreover in several case optical follow-up of the X-ray source is not 100\% complete,
therefore one must carefully reproduce the mask effect. What is usually done is that to 
create random samples in 3D, sources are placed at the same angular position of the 
real sources and redshift are randomly drawn from a smoothed redshift distribution 
of the real sources.  If instead the spectral completeness is close to 100\% then 
the right procedures is to occupy the survey volume with  random sources drawn 
from a L-z dependent luminosity function and accept check if they would be 
observable using a sensitivity map.
An  important choice for obtaining a reliable estimate of $w(r_{\rm p})$,
is to set $\pi_{max}$ in the calculation of the integral above. 
One should avoid values of  $\pi_{max}$ too large since they would  add noise to
the estimate of   $w(r_{\rm p})$.  If, instead,   $\pi_{max}$ is too small  one could not recover
all the signal. 
Uncertainties in the ACF are usually evaluated with a bootstrap resampling technique
but  it is worth noting that in the literature, several methods  
are adopted for errors estimates in two-point statistics \citet[for a detailed description]{Nor09}.
It is  known  that Poisson estimators generally underestimate the variance because 
they do consider that points in ACF are not  statistically independent. 
Jackknife resampling method, where one divides the survey area in many
sub fields and iteratively re-computes correlation functions by
excluding one sub-field at a time, generally gives a good estimates of 
errors. But it requires that sufficient number of almost statistically
independent sub- fields, this is not the case  for most of X-ray surveys where the
source statistics is moderately low. 
\citet{Coi09} estimated the error bars on the two-point correlation function including both Poisson 
and cosmic variance errors estimated, using DEEP2 mock 
catalogs derived from the Millenium Run simulations.

\section{Limber's Deprojection}

 The 2D Angular Correlation Function (ACF) is a projection of the real-space 3D ACF of the sources 
  along the line of sight. In the following discussions and thereafter,
   r is in comoving coordinates. The relation between the 2D (angular) ACF and the 3D ACF 
   is expressed by the Limber equation (e.g., Peebles 1980). Under the  assumption that the scale length
    of the clustering is much smaller than the distance to the object, this reduces to 
    \begin{equation}\begin{split}
   w(\theta)&N^2=\int \left( \frac{dN}{dZ}\right)^2 \\
   \int& \xi\left(\sqrt{[d_A(z)\theta^2)+l^{2}(1+z]}\right) \left(\frac{dl}{dz}\right)^{-1}dl~dz,
    \end{split}\end{equation}
where dA(z) is the angular distance, N is the total number of sources, and dN/dz is the 
redshift distribution (per z) of the sources. The redshift evolution
 of the 3D correlation function is customarily expressed by 

\begin{equation}
\xi(r,z)=\left(\frac{r}{r_0}\right)^{-\gamma}(1+z)^{-3-\epsilon+\gamma},
\end{equation}
where  $\epsilon$=-3 and  $\epsilon$=$\gamma$- 3 correspond to the case where the correlation length is constant in physical
 and comoving coordinates, respectively. In these notations, the zero-redshift
  3-D correlation length r$_0$ can be related to the
  angular correlation length $\theta_0$  by 
\begin{equation}\begin{split}
 &r_0^{\gamma}=(N^2/S)\theta_0^{\gamma-1},\\
 &S=H_{\gamma} \int \left( \frac{dN}{dZ} \right)^{2} \left[\frac{c~d\tau(z)}{dz}\right]^{-1}\\
  &d_A^{1-\gamma}(1+z)^{-3-\epsilon}dz,\\
 & H_{\gamma}=\frac{\Gamma[(\gamma-1)/2]\Gamma(1/2)}{]\Gamma(1/2)},
\end{split}\end{equation}

where $\tau$(z) is the look-back time. 
We also define the comoving correlation length 
\begin{equation}
r_0(\bar{z_{eff}})=r_0(1+\bar{z_{eff}})^{-3-\epsilon+\gamma},
\end{equation}
at the effective redshift $\bar{z_{eff}}$, which is the median redshift of the
 contribution to the angular correlation (the integrand of the second term).
An essential ingredient of the deprojection process is the redshift distribution of the sources
and when  individual redshifts are not available this is derived from integration of the luminosity function. 

\section{1-halo and 2-halo terms in the HOD formalism}

In the halo model approach, the two-point
correlation function  of AGN is the sum of two contributions: the first
term (\textit{1-halo term}) is due to the correlation between objects in the
same halo and the second term (\textit{2-halo term}) arises because of the
correlation between two distinct halos: 
\begin{equation}
w_{p}(r_p) = w_{p,1h}(r_p)+w_{p,2h}(r_p)
\end{equation}

Recent articles prefer to express $w = (1 + w_{1h}) + w_{2h}$ 
\citep{Tin05, Zhe05, Bla08}, instead of $w = w_{1h} + w_{2h}$, 
as used in older articles. This is because $1 + \xi$ represents 
a quantity that is proportional to the number of pairs $\propto [1+\xi_{1h}]+[1+\xi_{2h}]$.
In this new convention, the projected correlation function
$w_{p,1h}$ represents the projection of $1 + \xi_{1h}$
rather than $\xi_{1h}$.\\
Similarly, one express the power spectrum of the distribution 
of the AGN in terms of the 1- and 2-halo term contributions:
\begin{equation}
P (k) = P_{1h}(k) + P_{2h}(k)
\end{equation}
and then the projected correlation function as:
\begin{eqnarray}
w_{p,1h}(r_p) =\int \frac{k}{2\pi}P_{1h}(k)J_0(kr_p)dk \\
w_{p,2h}(r_p) = \int \frac{k}{2\pi} P_{2h}(k)J_0(kr_p)dk
\end{eqnarray}
where $J_0(x)$ is the zeroth-order Bessel function of the 
first kind. 

Several parameterizations exist in literature for representing the DMH profile
\citep{Coo02, Kno08, Sta09} and the \citet{Nav97} (NFW) profile is a popular choice.
If  $y(k,M_h)$ expresses the Fourier transform of the NFW profile of
the DMH with mass $M_h$, normalized such that volume integral 
up to the virial radius is unity, then the
 one-halo term of the power spectrum can be written as:
\begin{equation}
P_{1h}(k) =	\frac{1}{n_{AGN}^2} \int n(M_h)N(M_h)|y(k, M_h)|^2 dM_h
\end{equation}
Assuming the linear halo bias model \citep{Mo96}, the 
two-halo term of the power spectrum reduces to: 
\begin{equation}
P_{2h}(k) = P_{m}(k) \left[ \frac{1}{n_{AGN}}\int  n(M_h)b(M_h)y(k, M_h) dM_h \right]^2
\end{equation}
Since the clustering on large scales is dominated by the two-halo term, 
it is fairly insensitive to the assumption of AGN distribution inside 
the hosting halo \citep{Ber02}. It should be noted that 
since $y \sim 1$ on large scales (e.g. scales much larger than the 
virial radius of halos), on such scales the two-halo term can be rewritten as:
\begin{equation}
P_{2h}(k) \approx b^2P_{m}(k, z)
\end{equation}
or in terms of projected correlation function:
\begin{equation}\label{eq:wrp}
w_{p,2h}(r_p)=b^{2}w_{m,2h}(r_p)
\end{equation}
where $b$ is the bias parameter of the sample and $w_{m,2h}$ is the DM projected correlation function.
For the matter power spectrum, $P_{m}(k)$, one can use the primordial 
power spectrum with a fixed $n_s$ and a transfer function calculated 
using the fitting formula of \citet{Eis98} or the nonlinear
form given by \citet{Smith}, Tinker et al. (2005).

\section{Bias Parameter calculation}

In the majority of works on clustering of X-ray AGN \citep{Mul04, Yan06, Gil05, Coi09,
Kru10, Cap10} the standard approaches used to estimate the bias 
are based on the power-law fit parameters of the AGN correlation function.
This method assumes that the projected correlation function is well
fitted by a power-law and the bias factors are derived from
the best fit parameters $r_{0}$ and $\gamma$ of the clustering signal at large scale.
Using the power-law fit one can estimate
the AGN bias factor using the power-law best fit parameters:
\begin{equation}\label{eq:PLb}
b_{PL}=\sigma_{8,AGN}(z)/\sigma_{DM}(z)
\end{equation}
where $\sigma_{8,AGN}(z)$ is the rms fluctuations of the 
density distribution over the sphere with a comoving
radius of 8 Mpc $h^{-1}$, $\sigma_{DM}(z)$ is the dark matter correlation function 
evaluated at 8 Mpc $h^{-1}$, normalized to a value of $\sigma_{DM}(z=0)=0.8$. For a
power-law correlation function this value can be calculated by \citep{pee80}:
\begin{equation}\label{eq:bias}
(\sigma_{8,AGN})^{2} = J_2(\gamma)(\frac{r_0}{8 Mpc/h})^{\gamma}
\end{equation}
where $J_2(\gamma)=72/[(3-\gamma)(4-\gamma)(6-\gamma)2^{\gamma}]$.  

Differently in the halo model approach, 
the 2-halo term of the projected correlation function, which dominates at large scales, can
be considered in the regime of linear density fluctuations. 
In the linear regime, AGN are biased tracers of the dark matter
distribution and the bias factor is described by:
\begin{equation}\label{eq:b}
b=(w_{p,1h}(r_p)/w_{m,2h}(r_p))^{1/2}
\end{equation}
HOD modeling is currently the optimal method to establish the large-scale bias parameter, provided the 
parametrization of $N(M_h)$, by using:
\begin{equation}\label{bHOD}
b = \frac{\int b_h(M_h) N(M_h)n(M_h)dM_h}{\int N(M_h)n(M_h)dM_h}
\end{equation}
assuming the halo mass function $n(M_h)$and the halo bias factor $b(M_h)$.

In fact, power law fit bias measurements commonly use smaller 
scales ($<1-2h^{-1}$ Mpc) that are in the 1-halo term
in order to increase the statistical significance. If power law fits are restricted only to larger scales, the method suffers 
from the problem that the lowest scale, where the linear biasing 
scheme can still be applied, varies from sample to sample and remains ambiguous. \\
HOD modeling allows, in principle, the use of the full range of
 scales since the method first determines the 1 and 2-halo terms 
and then constrains the linear using data down to the smallest $r_p$ values
 that are dominated by the 2-halo term for each individual sample.\\
\citet{Kru11} estimated the RASS-AGN bias following the power-law
 (Eq. \ref{eq:PLb}) and the HOD (Eq. \ref{bHOD}) approach, pointing out that
using the first method the errors on the bias are much larger,
 but the values are statistically consistent which those derived from the HOD model fits.
\citet{Alle11} found similar results in estimating the COSMOS-AGN bias following Eq. \ref{eq:PLb} and \ref{eq:b}.
 
In order to derive a reliable picture of AGN clustering, bias parameters should be inferred from HOD modeling, 
or at least from the comparison of the correlation function with that of the DM only in the linear regime, because 
systematic errors based on power-law bias parameters will 
be larger than the statistical uncertainties of the clustering measurement.

\clearpage

\addtolength{\hoffset}{-1cm}
\begin{table}\label{tab1}
\tiny         
\centering
\begin{minipage}[t][180mm]{\textwidth}
\begin{tabular}{ccccccccc}
\hline
\hline
Survey  & Band &  N$_{obj}$ & z & $\theta_0$ & r$_{0}$ & $\gamma$ & b(z)$^{a}$ & Log(M$_{DMH}$)$^{b}$ \\
	      & keV &   &  & arcsec & $h^{-1}$ Mpc &  &   &    $\frac{M}{M_{\odot} h}$ 		\\
\hline
EMSS                & 0.5-2 &  183     &  $<$0.2   &        X                                          &       $<$10                    &   X                                   &    X                                     &  X   \\
RASS                &  0.1-2.4 & 2158   &  1-1.5      & $\sim$10                                     &       $<$10                     &  1.7$\pm{ 0.3}$             &    X                                     &  X   \\             
RASS                &  0.1-2.4 & 2096   &  0.1          & $\sim$3.7                                    &   6.0$\pm{1.6}$             &  1.9$\pm{0.31} $           &    X                                     &  X  \\   
ROSAT-NEP     &  0.1-2.4 &  220     &  0.22        &       X                                           &   7.5$^{+2.7}_{-4.2}$     &  1.85$^{+1.90}_{-0.80}$ & 1.83$^{+1.88}_{-0.61}$   &  13.51$^{+0.91}_{-0.79}$ \\   
AXIS$^{1}$                & 0.5-2 &  31288          &   0.96      & 22.9$\pm{2.0}$     &  6.54$\pm{0.12}$  & 1.12$\pm{0.04}$  &                 2.48$\pm{0.07}$      & 13.20$^{+0.11}_{-0.12}$          \\  
AXIS$^{1}$                & 2-10 &   9188          &   0.94    &  29.2$^{+5.1}_{-5.7}$ &  9.9$\pm{2.4}$ &  2.33$^{+0.10}_{?0.11}$ &   2.38$\pm{0.51}$ & 13.14$^{+0.28}_{-0.41}$ \\ 
AXIS$^{1}$                & 5-10 &   1259         &   0.77 &        40.9$^{+19.6}_{-29.3}$ &      5.1$\pm{4.1}$ &     1.47$^{+0.43}_{?0.57}$     &    2.14$\pm{1.88}$  &    13.17$^{+0.84}_{-2.44}$  \\ 
ELAIS-S1  & 0.5-2&  392 &   0.4  & 5.2$\pm3.8$ &   9.8$^{+2.7}_{-4.3}$          &  1.8                      &   X                                    &  X  \\    
ELAIS-S1  & 2-10 &    205   &   0.4  & 12.8$\pm$7.8 &   13.4 $^{+2.7}_{-4.3}$          &  1.8                      &   X                                    &  X  \\    
CDFS	            & 0.5-2    &  97       &  0.84          &        X                                          &   8.6$\pm{1.2}$             &  1.33$\pm{0.11}$          &  2.64$^{+0.29}_{-0.30}$       & 13.41$^{+0.55}_{-0.18}$\\
CDFN$^{2}$               & 0.5-2 & 164     &  0.96           &        X                                         &   4.2$\pm{0.4}$              & 1.42$\pm{0.07}$           & 1.87$^{+0.14}_{-0.16}$   &    12.73$^{+0.12}_{-0.17}$\\
XMM-2dF$^{3}$    &    0.5-2 & 432     &  1.2           & 10.8$\pm{1.9}$                         &   $\sim$16                      & 1.8                                 & 1.9-2.7                              & 12.5-13.1 \\
XMM-LSS     &    0.5-2 & 1130   &  0.7          & 6.3$\pm{3}$     &  6$\pm{3}$                     &  2.2$\pm{0.2}$                    &   X                                      & X \\
XMM-LSS       &  2-10 & 413   &  0.7          & 42$^{+7}_{-13}$     &  6$\pm{3}$                     &  3.1$^{+1.1}_{-0.5}$                     &   X                                      & X \\
CLASXS         &0.5-8   & 233    &  1.2           &         X                                         &  8.1$^{+1.2}_{-2.2}$      & 2.1$\pm0.5$        & 3.58$^{+2.49}_{-1.38}$  & 12.86$^{+0.61}_{-0.16}$ \\
CDFN $^4$      &   0.5-8   & 252    &  0.8            &         X                                       &  5.8$^{+1.0}_{-1.5}$      & 1.38$^{+0.12}_{-0.14}$ & 1.77$^{+0.80}_{-0.15}$&13.53$^{+0.63}_{-0.71}$\\
XMM-COSMOS$^{5}$  & 0.5-2 & 1037    &  1.1      & 2.9$\pm{0.6}$ &   11.8$\pm{1.1}$, &        1.8    &                  3.7$\pm0.3$                      &   13.6$\pm0.1$\\
XMM-COSMOS$^{5}$ & 2-4.5&  545    &   0.9   &  1.2$^{+1.1}_{-0.9}$  &    6.9$^{+2.2}_{-3.1}$,   &        1.8    &                  2.5$^{+0.7}_{-1.0}$              &   13.3$^{+0.3}_{-0.7}$\\
XMM-COSMOS$^{5}$ & 4.5-10  &  151    &  0.6      &  6.5$^{+3.0}_{-2.7}$  &   12.7$^{+2.3}_{-2.7}$  &        1.8    &                  3.8$^{+0.6}_{-0.8}$                         &   13.9$\pm0.2$\\
XMM-COSMOS$^{6}$ & 0.5-2 &         538     &  0.98       &         X                                        & 8.65$^{+0.41}_{-0.48}$ & 1.88$^{+0.06}_{-0.07}$   &   3.08$\pm0.14$     &  13.51$^{+0.05}_{-0.07}$ \\
XMM-COSMOS$^{7}$  & 0.5-2         &  593     &  1.21       &         X                                        & 7.12$^{+0.28}_{-0.18}$ &  1.81$^{+0.04}_{-0.03}$  & 2.71 $\pm 0.14$             &13.10$^{+0.06}_{-0.07}$\\
SWIFT-BAT    & 15-55  &  199     &  0.045     &         X                                        & 5.56$^{+0.49}_{-0.43}$ &  1.64$^{+0.07}_{-0.08}$   & 1.21$^{+0.06}_{-0.07}$  &13.15$^{+0.09}_{-0.13}$\\
AEGIS  & 0.5-2		      &  113     &   0.9        &         X                                        & 5.95$\pm 0.90$             &  1.66$\pm 0.22$              & 1.97$^{+0.26}_{-0.25}$  & 13.0$^{+0.1}_{-0.4}$ \\
AGES        &0.5-2        &  362     &   0.51      &         X                                        & 4.5$\pm 0.6$                 &  1.6$\pm 0.1$                  & 1.35 $^{+0.06}_{-0.07}$            &12.60$^{+0.1}_{-0.1}$\\
ROSAT+SDSS &0.1-2.4  &  1552    &  0.27      &         X                                        & 4.28$^{+0.44}_{-0.54}$  &  1.67$^{+0.13}_{-0.12}$  & 1.11$^{+0.10}_{-0.12}$  &12.58$^{+0.20}_{-0.33}$\\
XMM-LSS & 0.5-2 & 4360 & 1.1 & 3.2$\pm{0.5}$ & 7.2$\pm{0.8}$ & 1.93$\pm{0.03}$ & 2.7$\pm{0.3}$ & 13.2$\pm{0.3}$\\
XMM-LSS & 2-10 & 1712 & 1.0 & 9.9$\pm{0.4}$ & 10.1$\pm{0.9}$ & 1.98$\pm{0.04}$& 3.3$\pm{0.3}$ & 13.7$\pm{0.3}$  \\
\hline
\hline

\end{tabular}
\centering

X:Unconstrained or undetermined\\
$^{a}$: Bias factors converted to a common cosmology ($\Omega_{\Lambda}=0.7$, $\Omega_{m}=0.3$, $\sigma_{8}=0.8$)\\
$^{b}$: DMH masses estimated using \citet{van02} and \citet{She01}\\
$^{1}$: Ebrero et al. (2009), fit ID=2, assuming no redshift evolution of the correlation length\\
$^{2}$: Gilli et al. (2005)\\
$^{3}$: Basilakos et al. (2005), using the LDDE model\\
$^{4}$: Yang et al. (2006)\\
$^{5}$: Miyaji et al.(2007), fit ID=6 with integral constrain, assuming redshift evolution of the correlation length\\
$^{6}$: Gilli et al. (2009)\\
$^{7}$: Allevato et al. (2011)\\

\end{minipage}

\end{table}
\newpage

\newpage

\end{document}